\documentclass{naturefigs}

\usepackage{graphicx}
\usepackage{epstopdf}
\usepackage{verbatim}
\usepackage{color}

\title{Chemical Reaction of Ultracold Atoms and Ions in a Hybrid Trap}

\author{Wade G. Rellergert$^{1}$, Scott T. Sullivan$^{1}$, Svetlana Kotochigova$^{2}$, Alexander
Petrov$^{2}$, Kuang Chen$^{1}$, Steven J. Schowalter$^{1}$, and Eric R. Hudson$^{1}$}

\begin{document}

\maketitle

\begin{affiliations}
 \item Department of Physics and Astronomy, University of California, Los Angeles, California 90095, USA
 \item Department of Physics, Temple University, Philadelphia, Pennsylvania 19122-6082, USA
\end{affiliations}

\begin{abstract}

Interactions between cold ions and atoms have been proposed for use in implementing quantum gates\cite{Idziaszek2007}, probing quantum gases\cite{Sherkunov2009}, observing novel charge-transport dynamics\cite{Cote2000}, and sympathetically cooling atomic and molecular systems which cannot be laser cooled\cite{Smith2005,Hudson2009}. Furthermore, the chemistry between cold ions and atoms is foundational to issues in modern astrophysics, including the formation of stars, planets, and interstellar clouds\cite{Smith1992}, the diffuse interstellar bands\cite{Reddy2010}, and the post-recombination epoch of the early universe\cite{Stancil1996b}. However, as pointed out in refs 9 and 10, both experimental data and a theoretical description of the ion-atom interaction at low temperatures, reached in these modern atomic physics experiments and the interstellar environment, are still largely missing. Here we observe a chemical reaction between ultracold $^{174}$Yb$^+$ ions and  $^{40}$Ca atoms held in a hybrid trap. We measure, and theoretically reproduce, a chemical reaction rate constant of $ \rm \bf K =(2\pm1.3)\times10^{-10}~cm^{3}s^{-1}$ for $ \rm \bf 1~mK \leq T \leq 10~K$, four orders of magnitude higher than reported for other heteronuclear cases. We also offer a possible explanation for the apparent contradiction between typical theoretical predictions and measurements of the radiative association process in this and other systems.
\end{abstract}

Of particular interest in cold ion-atom physics are binary charge exchange (CEX) reactions, which limit the performance of the aforementioned proposals and are important inputs into the astrophysical models.   Binary CEX reactions occur via three distinct mechanisms: (i) non-radiative charge transfer (nRCT), $A^+ + B \rightarrow A + B^+,$ (ii) radiative charge transfer (RCT), $A^+ + B \rightarrow A + B^+ + \gamma$, and (iii) radiative association (RA), $A^+ + B \rightarrow (AB)^+ + \gamma$.  Due to the scarcity of reaction data and full quantum calculations, a number of approximation techniques for these three processes have been developed, \textit{e.g.} approximating all RA rate constants by 10$^{-14}~$cm$^3$s$^{-1}$(ref. 11), Landau-Zener theory\cite{Butler1980}, the Demkov coupling method\cite{Demkov1964}, and the semi-classical optical potential method\cite{Zygelman1988}.  These techniques, which have varying degrees of accuracy depending on the dominant CEX pathway, must be experimentally verified as demonstrated by a recent measurement where a factor of $\sim$100 discrepancy was found with experiment\cite{Woodall2007,Luca2002}.

Total CEX rate constants have been reported for two ultracold ion-atom species\cite{Grier2009,Zipkes2010a}. In the first system the CEX rate constant for the resonant, homonuclear case of Yb$^+$~+~Yb was measured by Grier~\textit{et al.} to be $\sim$6$\times$10$^{-10}$~cm$^3$s$^{-1}$(ref. 16).  On the other hand, a measurement of the rate constant for the non-resonant, heteronuclear case of Yb$^+$~+~Rb by Zipkes~\textit{et al.} yielded $\sim$3.5$\times$10$^{-14}$~cm$^3$s$^{-1}$(ref. 17); an experiment on Ba$^+$~+~Rb appears to have yielded similar results\cite{Schmid2010}.  These values agree well with conventional wisdom for resonant versus non-resonant CEX.  Interestingly, although the RA process is typically predicted to be the dominant mechanism for low-temperature, non-resonant CEX\cite{Zygelman1989,Liu2009,Zhao2004}, none of these experiments observed the formation of a molecular ion.  Further, Zipkes~\textit{et al.} found the ratio of the rate of nRCT to RCT to be K$_{nRCT}$/K$_{RCT}\approx$ 2.3, while most calculations predict K$_{nRCT}$/K$_{RCT}\leq$ 10$^{-3}$ in the ultracold regime\cite{Zygelman1989,Liu2009,Zhao2004}.

Here we report a measurement of the CEX chemical reaction rate constant, K, for the non-resonant, heteronuclear case of $^{174}$Yb$^+$~+~$^{40}$Ca in a hybrid trap which is four orders of magnitude higher than reported for other heteronuclear cases. In addition, by monitoring the production of CaYb$^+$ molecular ions in the hybrid trap, we place an apparent upper bound on the branching ratio of the RA process. Molecular potential curves for the CaYb$^+$ molecule have also been constructed and used in a calculation that explains this enhanced rate constant as the result of an avoided crossing and strong transition dipole moment. In what follows, we describe the experimental setup, results, and theoretical model.  We conclude with a possible explanation for the apparent contradiction between typical theoretical predictions and measurements of the RA process in this and other experiments\cite{Zipkes2010a}.

\begin{figure}
\resizebox{1\columnwidth}{!}{
    \includegraphics{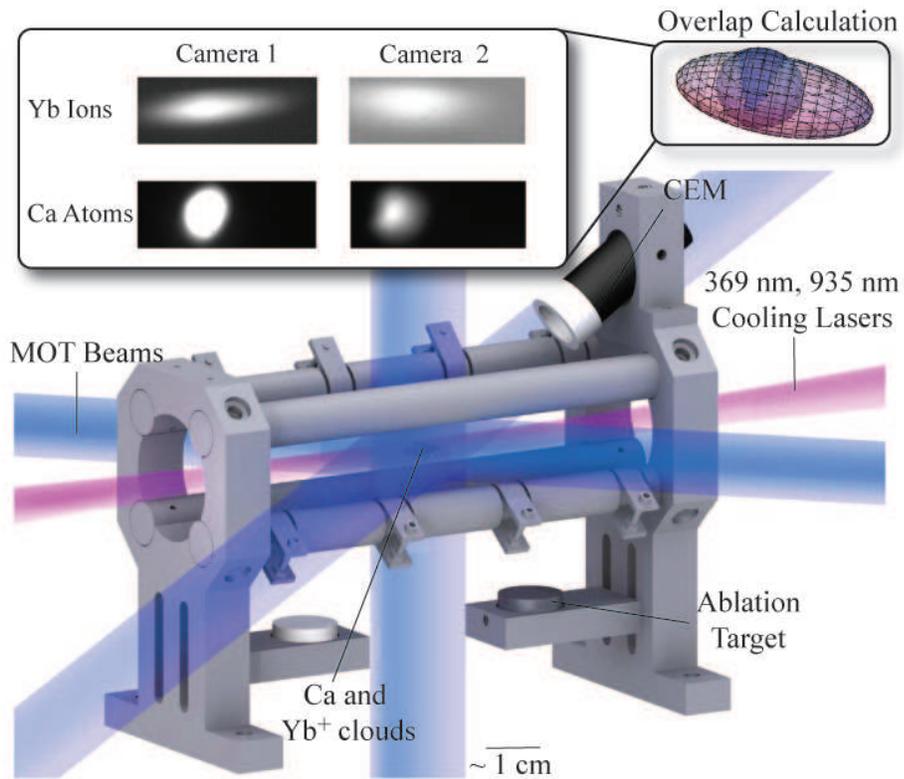}
}
\caption{MOTION trap schematic. Yb$^+$ ions and Ca atoms are laser cooled and trapped in the central region. Typical images of the ultracold clouds, shown in the upper left of the figure, are used for a 3D reconstruction that yields the degree of overlap.  Ions are also detected with a voltage-gated CEM.}
\label{setup}
\end{figure}

Figure~1 shows the hybrid magneto-optical and ion trap (MOTION trap) system used in these experiments\cite{Hudson2009}.  Yb$^{+}$ ions, created by laser ablation of a solid Yb target, are trapped in a linear quadrupole trap (LQT)~\cite{Douglas2005} and laser-cooled using laser beams ($\lambda$~=~369~nm,~935~nm) aligned along the trap axis.
In the same region, ultracold Ca atoms are produced and held in a magneto-optical trap (MOT) constructed with laser beams ($\lambda$~=~422~nm,~672~nm) that intersect at the LQT center (See Methods). Fluorescence images of both the Ca MOT and the ion cloud from two separate, nearly orthogonal camera angles allow a 3D  reconstruction of the ion and atom clouds and the degree of overlap is quantified as $\Phi = \int \hat{\rho}_{Ca} (\vec{r}) \bar{\rho}_{I}(\vec{r}) d\vec{r} ~$, where $\hat{\rho}_{Ca}$ is the unit-peak normalized Ca atom density and $\bar{\rho}_{I}$ is the unit-integral normalized Yb$^+$ density (See methods).  An example of such images and the reconstruction is shown in the inset of  Fig.~1.

In absence of the Ca MOT, the trap lifetime of the laser-cooled Yb$^{+}$ ions, measured by monitoring the ion fluorescence, is $\sim$~240~s.  However, when an overlapped MOT is introduced, the ion lifetime decreases to $\sim$~10~s, as shown in Fig.~2a.  This induced ion loss is also confirmed visually by imaging the ion cloud during the decay, as well as observing a decreased number of detected ions using the CEM. The inset in Fig.~2a shows that the loss of ions is not due to ion heating, as such an effect would increase the ion cloud size.  From data such as these we calculate a loss rate constant using K~$ = \langle \sigma v \rangle = \frac{\Gamma }{{\rho} \Phi} $, where $\Gamma$ is the measured fluorescence decay rate and $\rho$ is the peak density of the MOT.

\begin{figure}
\resizebox{1\columnwidth}{!}{
  \includegraphics{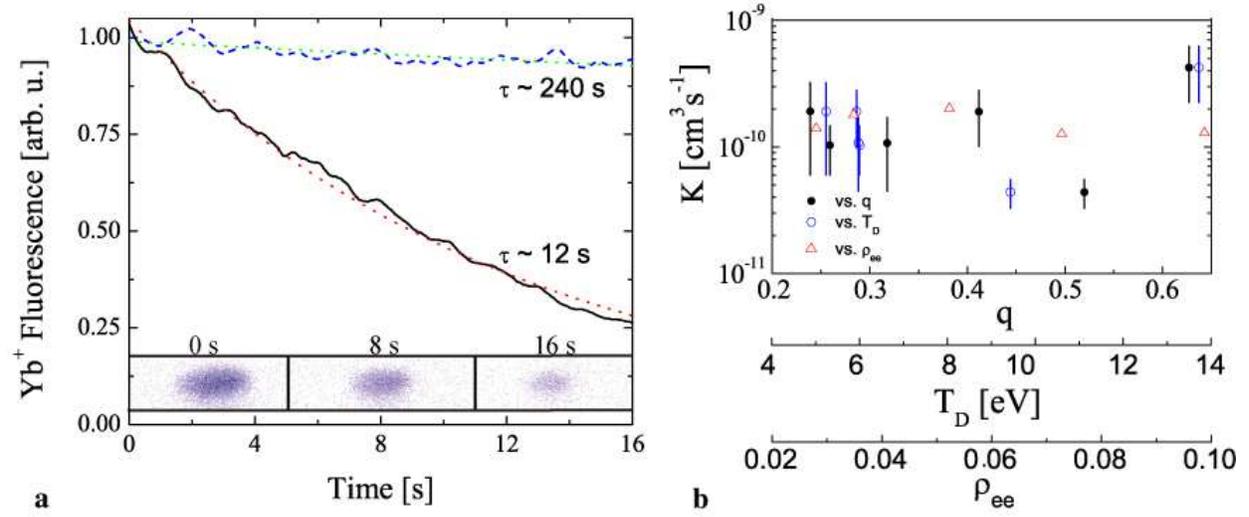}
}
\caption{\textbf{a} Yb$^+$ ion trap lifetime in the presence of the MOT (solid curve) compared to the lifetime without a MOT (dashed curve). Dotted curves are single exponential decay fits. Inset: Images of the Yb$^+$ ion cloud at various times. \textbf{b} Loss rate constant as a function of the Mathieu stability parameter q,  the trap depth T$_D$, and the excited state fraction of Ca, $\rho_{ee}$.  }
\label{decay_curve_and_rates}
\end{figure}

Given the large measured rate constant and the disparity between previous work and typical theoretical predictions, we performed several experiments to identify possible systematic errors. First, in all previous ultracold experiments the rate constant was measured, as it was here, by observing the decay of the parent ion, rather than the rate of appearance of the reaction products, which are often untrappable. Thus, we set out to verify that no kinematic effects were responsible for the loss of Yb$^+$ ions from the trap.  To this end, we measured the loss rate constant while varying only q, the Mathieu parameter, which determines the ions' stability and micromotion amplitude in the LQT\cite{Douglas2005}, as well as when only varying the trap depth, T$_D$.  These data are shown in Fig.~2b.  As the rate constant appears independent of these fundamental trapping parameters, it is unlikely that the loss is due to a kinematic effect in the LQT.

In addition, we performed an experiment where BaCl$^+$ ions, produced by laser ablation of an additional target (see Fig.~1), were sympathetically cooled by co-trapped, laser cooled Yb$^+$ ions.  We then compared, after 20~s of interaction with the MOT, the number of ions remaining in the trap and found that while the Yb$^+$ ions had decayed away, the BaCl$^+$ ions remained. As BaCl$^+$ ions are energetically forbidden to charge exchange with Ca atoms\cite{Hudson2009}, we conclude the loss of Yb$^+$ ions is due to a CEX reaction.

Finally, as discussed in ref. 23, it is expected that the observed reaction is between ground state atoms and ions, since the long-range ion-neutral interaction and ultracold temperatures preclude excited state atoms and ions from being in close enough proximity to undergo a chemical reaction. To verify this, the loss rate constant was measured as a function of the excited state fraction, $\rho_{ee}$, of Ca atoms in the MOT, by varying the cooling laser intensities. As seen in Fig.~2b, no dependence was observed and we conclude that the loss of Yb$^+$ fluorescence is due to a CEX chemical reaction of ground state Ca atoms and Yb$^{+}$ ions.

\begin{figure}
\resizebox{1\columnwidth}{!}{
  \includegraphics{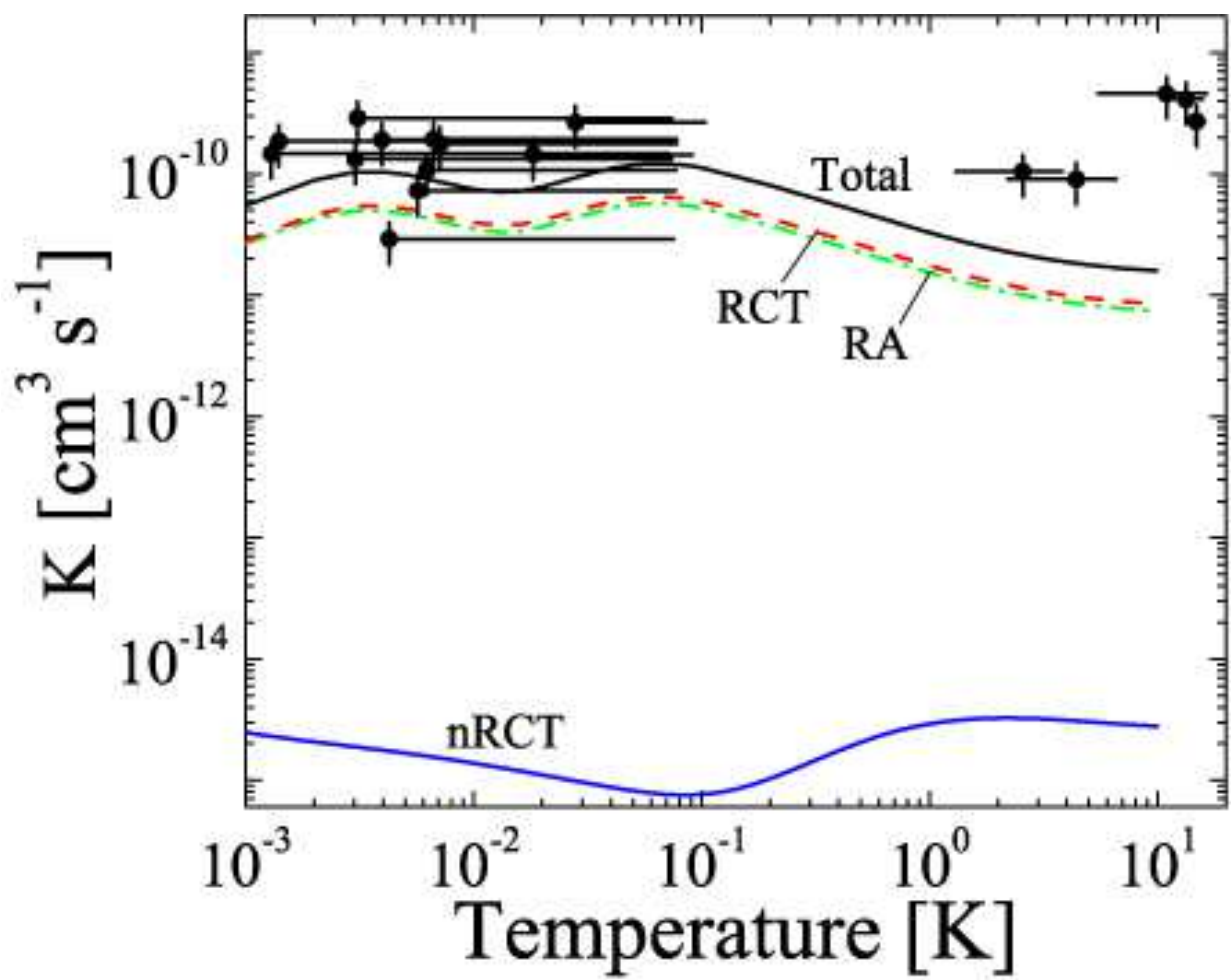}
}
\caption{Total experimental (points) and theoretical (black line) charge exchange rate constants. Also shown are the calculated rate constants for the three pathways. }
\label{ratevstemp}
\end{figure}

After these systematic checks the loss rate constant was measured as a function of temperature, as shown in Fig.~3.  For these measurements the MOT temperature was held at 4~mK and the ion temperature varied by changing the cooling laser detuning and intensity. The horizontal error bars account for the possible contribution to the average ion energy that results from residual radial micromotion\cite{Major1968}.  The observed rate constant is roughly four orders of magnitude larger than typical non-resonant, heteronuclear CEX values and is constant within experimental error over all measured temperatures.

The CEX reaction products can in principle be trapped, but due to the large mass ratio of Ca$^+$ (m~=~40 amu) and Yb$^+$ (m~=~174 amu) ions, Ca$^+$ ions are heated by the cold Yb$^+$ ions\cite{Major1968}. This effect is exacerbated by the fact that trap operation in a mutually stable region necessarily puts one or both ion species near a boundary of their LQT stability region. As a result, at the trap parameters used in this study, Ca$^+$ ions have a measured trap lifetime of only $\sim$1~s.  Given the comparatively low production rate from CEX reactions with Yb$^{+}$ ions, the equilibrium number of Ca$^+$ ions in the trap is $\leq$~2, which is below our current detection limit. Conversely, the mass of the CaYb$^{+}$ reaction product is much closer to that of Yb$^+$ and does not suffer from these effects. Therefore, neglecting further chemical reaction with the Ca MOT, CaYb$^+$ ions are expected to have a trap lifetime of at least that of uncooled BaCl$^{+}$ ions (m~$\sim$~174 amu) which we measure to be $\sim$30~s. Thus, by allowing the Yb$^+$ ion fluorescence to decay to background levels and then measuring the number of ions remaining in the trap using the CEM, we place an apparent upper limit on the branching ratio for RA at 0.02, under the assumption that all detected ions are CaYb$^+$.

Also shown in Fig.~3 are the results of our theoretical model for the CEX chemical reaction rate constant. Since the structure of CaYb$^+$ was not previously known, we first calculated the electronic potential surfaces that dissociate to the Ca+Yb$^+$ and Ca$^+$+Yb limits, as well as their transition dipole moment using the MOLCAS software suite\cite{Karlstrom2003}, as shown in Fig.~4a~and~b~(solid curves), respectively.  This approach worked well for the structure of the BaCl$^+$ molecular ion\cite{Chen2011}.

\begin{figure}
\resizebox{1\columnwidth}{!}{
\includegraphics{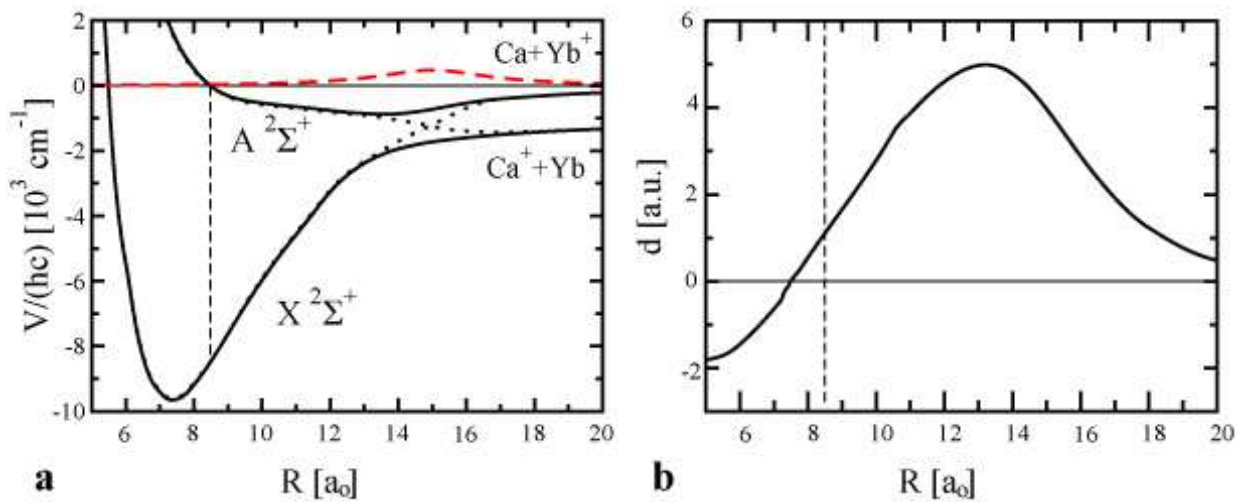}
}
\caption{\textbf{a} The  X $^2\Sigma^+$ and A$^2\Sigma^+$  potential 
energy curves and \textbf{b} the corresponding transition dipole moment of the CaYb$^+$ molecular ion as a function of the internuclear
separation $R$. The dotted curves are the diabatized potentials
constructed from these potentials. The dashed  curve gives the
coupling constant between the diabatic potentials. The vertical dotted line shows the inner turning point at zero collision energy of the A$^2\Sigma^+$  potential and the corresponding transition dipole moment.
}
\label{structure}
\end{figure}

Using these potentials, we calculated the rate constant for the CEX pathways via a quantum mechanical scattering calculation that assumed the particles collide on the A$^2\Sigma^+$ potential and charge exchange to bound or continuum states of the X$^2\Sigma^+$ potential.  Since the nRCT pathway occurs via a nonadiabatic transition near the avoided crossing, we determine its rate constant by a coupled-channels method with diabatized potentials -- black dotted lines in Fig.~4a (See Methods).  The thermalized nRCT rate constants are shown as a function of temperature in Fig.~3.  We have  varied the strength of the coupling between the diabatic potentials by 20\% to characterize the nature of the coupling.  Although we find that the nRCT rate coefficient dramatically increases with increasing coupling strength, it does not exceed a few times  $10^{-14}$~cm$^3$s$^{-1}$, indicating that the scattering is almost adiabatic.

For the calculation of the RCT and RA rate constants, we use the perturbative Fermi's golden rule for the spontaneous emission of a photon by the diatom (See Methods). Emission to the continuum of the X$^2\Sigma^+$ potential is characterized as RCT, while emission to a bound state of the potential is characterized as RA. Figure~3 shows the thermalized radiative rate constants due to both RCT and RA mechanisms versus temperature; in both cases on the order of a hundred partial waves are included to ensure numerical convergence. We find rate coefficients that are many times $10^{-11}$ cm$^3$/s and also observe that CEX should produce nearly equal amounts of Ca$^{+}$ atomic ions and CaYb$^{+}$ molecular ions.

While the total calculated rate constant, also shown in Fig.~3, exhibits satisfactory agreement with the total measured rate constant, the predicted RA branching ratio of $\sim 0.5$ strongly disagrees with the measured value $\leq$~0.02. These ratios could be reconciled, however, if the CaYb$^+$ further reacted with the Ca MOT atoms via subsequent RA reactions, \textit{e.g.}~CaYb$^+$~+~Ca~$\rightarrow$~Ca$_2$Yb$^+$~+~$\gamma$, to form heavier molecular ions, which would be unstable in our LQT. Given that RA will predominantly produce CaYb$^+$ in high-lying vibrational states, which are typically very reactive, this explanation seems quite plausible. Further, if the same explanation is applied to the data of ref. 1, the interpretation of trap loss as evidence of nRCT could be reinterpreted as evidence of RA, yielding a RA branching ratio that agrees well with typical theoretical predictions.

In conclusion, we have presented an experimental and theoretical investigation of the CEX reaction of the Ca + Yb$^+$ system. The total measured and calculated rate constants are in good agreement and are surprisingly large for a non-resonant, heteronuclear CEX reaction. We have also offered a plausible explanation for the apparent contradiction between typical theoretical predictions and measured rates of RA in this and other experiments. Given the importance of these types of reactions for determining astrophysical processes, this work highlights the need for fully quantum calculations of atom-ion systems and a renewed effort in laboratory astrophysics.  Finally, we note that although the CEX rate constant of this system is large, it should not interfere with most of the proposed applications of hybrid atom-ion systems.

\begin{methods}
\subsection{MOTION Trap.}

The MOTION trap (Fig.~1) consists of a radio-frequency linear quadrupole trap (LQT)~\cite{Douglas2005} (250~kHz~$\leq$~f~$\leq$~400~kHz, V$_{\rm{RF}}$ $\approx$ 300~V) with four sets of
segmenting electrodes to provide axial confinement. A channel electron multiplier (CEM) mounted above the trap provides a species non-specific method of ion detection, and is voltage-gated so that it does not affect the ion dynamics before detection.  Yb$^{+}$ ions, created by laser ablation of a solid Yb target
with a focused $\sim$1~mJ, 10~ns, 1064~nm laser pulse, are trapped in the middle region of the trap and laser-cooled using two laser beams aligned along the trap axis; one with a wavelength and intensity of $\lambda$~=~369~nm and $I\approx$~400~mW/cm$^2$, respectively, to excite the cooling transitions, and one with $\lambda$~=~935~nm and $I\approx$~2000~mW/cm$^2$ to regain ions which fall out of the cooling cycle.

The ultracold Ca atoms are produced and held in a magneto-optical trap (MOT) using 6 primary trapping laser beams ($\lambda=422$~nm, $I\approx4$~mW/cm$^2$), a deceleration beam ($\lambda=422$~nm, $I\approx31$~mW/cm$^2$), and a repumping beam ($\lambda=672$~nm, $I\approx 70$~mW/cm$^2$), all of which intersect at the LQT center.  The source of Ca is a chromate-free getter, $\sim$5~cm from the MOT center.  The typical MOT atom number, density, and temperature are measured by absorption and fluorescence imaging and found to be $N_{\rm{Ca}}$ = $3.7\pm 0.5\times10^6$~atoms, $\rho_{\rm{Ca}}$ = $7 \pm 1 \times 10^{9}$~cm$^{-3}$, and $T_{\rm{Ca}}$ = $4 \pm 1$~mK, respectively.

Ion number and temperature are determined by observing ion fluorescence while scanning the 369~nm laser ($\dot{f} =  500$~MHz/s) and fitting the resultant truncated Voigt profile\cite{Kielpinski2006}.  As the ion temperature is known, this allows for a determination of the total number of ions in the trap giving, among other things, a method to calibrate the CEM.  The laser is then slowly scanned up to a specific, predetermined red-detuned frequency and is locked via a transfer cavity to the 422~nm laser that is referenced to the Ca 4$^1$P$\leftarrow$4$^1$S transition.

When operating the LQT in a region where both Ca$^+$ and Yb$^+$ ions are stable,
a large heat load on the laser cooled Yb$^+$ ions is observed due to the ionization of Ca atoms in the 4$^1$P state by the Yb$^+$ cooling laser. To eliminate this effect, we modulate the intensity of the 422~nm and 369~nm lasers out of phase.  Additionally, this effect is eliminated if the LQT is operated where Ca$^+$ ions are not trapped.

\subsection{Alignment of MOT and Ion Clouds and Overlap Calculation.}
The position of the ions can be shifted axially in the trap by applying an asymmetric bias to the DC segmenting electrodes, and the location of the Ca MOT center can be changed both by adjusting the magnetic field and relative laser intensities.  These controls allow the ion and atom cloud positions to be precisely overlapped, which is the starting point for the data reported here.  We  quantify this overlap, in order to determine the effective density ($\rho \Phi$) of neutral Ca, by imaging both clouds with two, nearly orthogonal cameras.  The images, along with the known distances and angles of the cameras, allow a full 3D projection of cloud density distributions in the lab reference frame.  We then determine the overlap factor by convolving the unit-peak normalized MOT density distribution with the unit-integral normalized ion cloud distribution.  In addition, we time average the overlap factor by taking into account  the diminishing ion cloud size in coincidence with the fluorescence decay (see Fig.~2a).

\subsection{nRCT Calculation.}

We determine the nRCT rate constant by a coupled-channels method with diabatized potentials. For the coupling between these  potentials, we assume a Lorentzian in $R$ with a maximum value equal to half the minimum splitting of the adiabatic potentials. The width of the coupling matrix element is chosen such that diagonalization of the 2$\times$2 potential matrix reproduces the original adiabatic potentials.
Due to the strength of the atom-ion interaction, up to several hundred partial waves are required for numerical convergence of the nRCT rate constants at our temperatures, which leads to a large number of shape resonances in the unthermalized rate constants as a function of collision energy.

\subsection{RCT and RA Calculation.}

For the calculation of the RCT and RA rate constants, we use the perturbative Fermi's golden rule for the spontaneous emission of a photon by the diatom.
In principle, both the initial and final state wavefunctions are solutions of a two-channel calculation as for nRCT, however, since the scattering is nearly adiabatic, we assume that the initial state is a scattering solution of the A$^2\Sigma^+$ potential and the final state is either a bound or scattering state of the X $^2\Sigma^+$ potential. Thus, the total radiative rate coefficient for an initial Ca+Yb$^+$ energy-normalized state $|{\rm A}, \epsilon \ell m \rangle$ with collision energy $\epsilon$, relative angular momentum quantum number $\ell$, and projection $m$ is
\begin{eqnarray}
K_{\epsilon \ell m} = \frac{2\pi^2}{\sqrt{2\mu \epsilon}}   \sum_{f}
 \frac{4}{3} \alpha^3 \omega_{f,\epsilon \ell}^3
                |\langle {\rm X}, f | d |{\rm A}, \epsilon \ell m \rangle|^2\,
\label{spont}
\end{eqnarray}
where $\alpha$ is the fine-structure constant, $\mu$ is the reduced mass, and all quantities are expressed in atomic units. The sum is over final states $f$, which denote either a continuum wavefunction $|{\rm X}, \epsilon' \ell'm'\rangle$ leading to RCT or a ro-vibrational state $|{\rm X}, v\ell' m'\rangle$ of the X$^2\Sigma^+$ potential leading to RA. The quantity  $\omega_{f,\epsilon \ell}$ is the frequency difference between the initial and final states and the operator $d$ is an abbreviation for $d(R)C_{1q}(\hat R)$, where $d(R)$ is the $R$-dependent dipole moment between the X and A state, $C_{1q}(\hat R)$ is a spherical harmonic, $\hat{R}$ the orientation of the molecule in a space-fixed coordinate system, and $q=m-m'$ gives the polarization of the emitted photon. The selection rules of the dipole moment ensure that $|\ell-\ell'|\le 1$ and $|m-m'|\le 1$.

\end{methods}

\begin{addendum}
\item This work was supported by NSF grant No. PHY-1005453, ARO grant No. W911NF-10-1-0505 and AFOSR grant. ERH thanks Phillip Stancil for illuminating discussions.

\item[Competing Interests] The authors declare that they have no
competing financial interests.

\item[Correspondence] Correspondence and requests for materials
should be addressed to \hfill \ \linebreak  E.R.H.~(email: eric.hudson@ucla.edu).
\end{addendum}



\begin{thebibliography}{10}
\expandafter\ifx\csname url\endcsname\relax
  \def\url#1{\texttt{#1}}\fi
\expandafter\ifx\csname urlprefix\endcsname\relax\def\urlprefix{URL }\fi
\providecommand{\bibinfo}[2]{#2}
\providecommand{\eprint}[2][]{\url{#2}}

\bibitem{Idziaszek2007}
\bibinfo{author}{Idziaszek, Z.} \bibinfo{author}{Calarco, T.} \&
  \bibinfo{author}{Zoller, P.}
\newblock \bibinfo{title}{{Controlled collisions of a single atom and an ion
  guided by movable trapping potentials}}.
\newblock \emph{\bibinfo{journal}{Phys. Rev. A}}
  \textbf{\bibinfo{volume}{76}}, \bibinfo{pages}{1--16} (\bibinfo{year}{2007}).


\bibitem{Sherkunov2009}
\bibinfo{author}{Sherkunov, Y.}, \bibinfo{author}{Muzykantskii, B.},
  \bibinfo{author}{D’Ambrumenil, N.} \& \bibinfo{author}{Simons, B.}
\newblock \bibinfo{title}{{Probing ultracold Fermi atoms with a single ion}}.
\newblock \emph{\bibinfo{journal}{Phys. Rev. A}}
  \textbf{\bibinfo{volume}{79}}, \bibinfo{pages}{1--5} (\bibinfo{year}{2009}).


\bibitem{Cote2000}
\bibinfo{author}{C\^{o}t\'{e}, R.}
\newblock \bibinfo{title}{{From classical mobility to hopping conductivity:
  charge hopping in an ultracold gas.}}
\newblock \emph{\bibinfo{journal}{Phys. Rev. Lett.}}
  \textbf{\bibinfo{volume}{85}}, \bibinfo{pages}{5316--9}
  (\bibinfo{year}{2000}).


\bibitem{Smith2005}
\bibinfo{author}{Smith, W.~W.}, \bibinfo{author}{Makarov, O.~P.} \&
  \bibinfo{author}{Lin, J.}
\newblock \bibinfo{title}{{Cold ion-neutral collisions in a hybrid trap}}.
\newblock \emph{\bibinfo{journal}{J. Mod. Opt.}}
  \textbf{\bibinfo{volume}{52}}, \bibinfo{pages}{2253--2260}
  (\bibinfo{year}{2005}).
\newblock


\bibitem{Hudson2009}
\bibinfo{author}{Hudson, E.}
\newblock \bibinfo{title}{{Method for producing ultracold molecular ions}}.
\newblock \emph{\bibinfo{journal}{Phys. Rev. A}}
  \textbf{\bibinfo{volume}{79}}, \bibinfo{pages}{1--9} (\bibinfo{year}{2009}).


\bibitem{Smith1992}
\bibinfo{author}{Smith, D.}
\newblock \bibinfo{title}{{The ion chemistry of interstellar clouds}}.
\newblock \emph{\bibinfo{journal}{Chem. Rev.}}
  \textbf{\bibinfo{volume}{92}}, \bibinfo{pages}{1473--1485} (\bibinfo{year}{1992}).

\bibitem{Reddy2010}
\bibinfo{author}{Reddy, V.~S.}, \bibinfo{author}{Ghanta, S.} \&
  \bibinfo{author}{Mahapatra, S.}
\newblock \bibinfo{title}{First principles quantum dynamical investigation
  provides evidence for the role of polycyclic aromatic hydrocarbon radical
  cations in interstellar physics}.
\newblock \emph{\bibinfo{journal}{Phys. Rev. Lett.}}
  \textbf{\bibinfo{volume}{104}}, \bibinfo{pages}{111102}
  (\bibinfo{year}{2010}).

\bibitem{Stancil1996b}
\bibinfo{author}{Stancil, P.} \& \bibinfo{author}{Zygelman, B.}
\newblock \bibinfo{title}{Radiative charge transfer in collisions of Li with
  H$^+$}.
\newblock \emph{\bibinfo{journal}{Astrophys. J.}}
  \textbf{\bibinfo{volume}{472}}, \bibinfo{pages}{102--107}
  (\bibinfo{year}{1996}).

\bibitem{Idziaszek2009}
\bibinfo{author}{Idziaszek, Z.}, \bibinfo{author}{Calarco, T.},
  \bibinfo{author}{Julienne, P.} \& \bibinfo{author}{Simoni, A.}
\newblock \bibinfo{title}{{Quantum theory of ultracold atom-ion collisions}}.
\newblock \emph{\bibinfo{journal}{Phys. Rev. A}}
  \textbf{\bibinfo{volume}{79}}, \bibinfo{pages}{1--4} (\bibinfo{year}{2009}).

\bibitem{Woodall2007}
\bibinfo{author}{Woodall, J.}, \bibinfo{author}{Agundez, M.},
  \bibinfo{author}{Markwick-Kemper, A.~J.} \& \bibinfo{author}{Millar, T.~J.}
\newblock \bibinfo{title}{The UMIST database for Astrochemistry 2006}
  (\bibinfo{year}{2007}).

\bibitem{Kingdon1995}
\bibinfo{author}{Kingdon, J.}
\newblock \bibinfo{title}{Landau-Zener charge exchange}.
\newblock \emph{\bibinfo{journal}{Mon. Not. R. Astron. Soc.}}
  \textbf{\bibinfo{volume}{274}}, \bibinfo{pages}{425} (\bibinfo{year}{1995}).

\bibitem{Butler1980}
\bibinfo{author}{Butler, S.~E.} \& \bibinfo{author}{Dalgarno, A.}
\newblock \bibinfo{title}{Charge transfer of multiply charged ions with
  hydrogen and helium Landau-Zener calculations}.
\newblock \emph{\bibinfo{journal}{Astrophys. J.}}
  \textbf{\bibinfo{volume}{241}}, \bibinfo{pages}{838} (\bibinfo{year}{1980}).

\bibitem{Demkov1964}
\bibinfo{author}{Demkov, Y.~N.}
\newblock \bibinfo{title}{Charge transfer at small resonance defects}.
\newblock \emph{\bibinfo{journal}{Sov. Phys. JETP}}
  \textbf{\bibinfo{volume}{18}}, \bibinfo{pages}{138} (\bibinfo{year}{1964}).

\bibitem{Zygelman1988}
\bibinfo{author}{Zygelman, B.} \& \bibinfo{author}{Dalgarno, A.}
\newblock \bibinfo{title}{Radiative quenching of He(2 $^1$S) induced by
  collisions with ground-state helium-atoms}.
\newblock \emph{\bibinfo{journal}{Phys. Rev. A}} \textbf{\bibinfo{volume}{38}},
  \bibinfo{pages}{1877--1884} (\bibinfo{year}{1988}).



\bibitem{Luca2002}
\bibinfo{author}{Luca, A.}, \bibinfo{author}{Voulot, D.} \&
  \bibinfo{author}{Gerlich, D.}
\newblock \bibinfo{title}{Low temperature reactions between stored ions and
  condensable gases: formation of protonated methanol via radiative
  association}.
\newblock \emph{\bibinfo{journal}{WDS Proc. of Contrib. Pap.,}}
  \textbf{\bibinfo{volume}{2}}, \bibinfo{pages}{294} (\bibinfo{year}{2002}).




\bibitem{Grier2009}
\bibinfo{author}{Grier, A.}, \bibinfo{author}{Cetina, M.},
  \bibinfo{author}{Oru\v{c}evi\'{c}, F.} \& \bibinfo{author}{Vuleti\'{c}, V.}
\newblock \bibinfo{title}{{Observation of cold collisions between trapped ions
  and trapped atoms}}.
\newblock \emph{\bibinfo{journal}{Phys. Rev. Lett.}}
  \textbf{\bibinfo{volume}{102}}, \bibinfo{pages}{1--4} (\bibinfo{year}{2009}).



\bibitem{Zipkes2010a}
\bibinfo{author}{Zipkes, C.}, \bibinfo{author}{Palzer, S.},
  \bibinfo{author}{Ratschbacher, L.}, \bibinfo{author}{Sias, C.} \&
  \bibinfo{author}{K\"{o}hl, M.}
\newblock \bibinfo{title}{{Cold heteronuclear atom-ion collisions}}.
\newblock \emph{\bibinfo{journal}{Phys. Rev. Lett.}}
  \textbf{\bibinfo{volume}{105}}, \bibinfo{pages}{1--4} (\bibinfo{year}{2010}).


\bibitem{Schmid2010}
\bibinfo{author}{Schmid, S.}, \bibinfo{author}{H\"{a}rter, A.} \&
  \bibinfo{author}{Denschlag, J.}
\newblock \bibinfo{title}{{Dynamics of a cold trapped ion in a Bose-Einstein
  condensate}}.
\newblock \emph{\bibinfo{journal}{Phys. Rev. Lett.}}
  \textbf{\bibinfo{volume}{105}}, \bibinfo{pages}{1--4} (\bibinfo{year}{2010}).


\bibitem{Zygelman1989}
\bibinfo{author}{Zygelman, B.}, \bibinfo{author}{Dalgarno, A.},
  \bibinfo{author}{Kimura, M.} \& \bibinfo{author}{Lane, N.~F.}
\newblock \bibinfo{title}{{Radiative and nonradiative charge transfer in He$^+$+H collisions at low energy}}.
\newblock \emph{\bibinfo{journal}{Phys. Rev. A}}
  \textbf{\bibinfo{volume}{40}}, \bibinfo{pages}{2340--2345}
  (\bibinfo{year}{1989}).

\bibitem{Liu2009}
\bibinfo{author}{Liu, C.} \emph{et~al.}
\newblock \bibinfo{title}{{Radiative charge transfer in collisions of H+ with
  Na at very low energies}}.
\newblock \emph{\bibinfo{journal}{Phys. Rev. A}}
  \textbf{\bibinfo{volume}{79}}, \bibinfo{pages}{1--5} (\bibinfo{year}{2009}).


\bibitem{Zhao2004}
\bibinfo{author}{Zhao, L.~B.} \emph{et~al.}
\newblock \bibinfo{title}{{Radiative charge transfer in collisions of O with He$^+$}}.
\newblock \emph{\bibinfo{journal}{Astrophys. J.}}
  \textbf{\bibinfo{volume}{615}}, \bibinfo{pages}{1063--1072}
  (\bibinfo{year}{2004}).


\bibitem{Douglas2005}
\bibinfo{author}{Douglas, D.~J.}, \bibinfo{author}{Frank, A.~J.} \&
  \bibinfo{author}{Mao, D.}
\newblock \bibinfo{title}{{Linear ion traps in mass spectrometry.}}
\newblock \emph{\bibinfo{journal}{Mass Spectrom. Rev.}}
  \textbf{\bibinfo{volume}{24}}, \bibinfo{pages}{1--29} (\bibinfo{year}{2005}).


\bibitem{Band1992}
\bibinfo{author}{Band, Y.~B.} \& \bibinfo{author}{Julienne, P.~S.}
\newblock \bibinfo{title}{Optical-bloch-equation method for cold-atom
  collisions: Cs loss from optical traps}.
\newblock \emph{\bibinfo{journal}{Phys. Rev. A}} \textbf{\bibinfo{volume}{46}},
  \bibinfo{pages}{330} (\bibinfo{year}{1992}).

\bibitem{Major1968}
\bibinfo{author}{Major, F.~G.} \& \bibinfo{author}{Dehmelt, H.~G.}
\newblock \bibinfo{title}{Exchange-collision technique for the RF spectroscopy of stored ions}.
\newblock \emph{\bibinfo{journal}{Physical Review}} \bibinfo{pages}{91}
  (\bibinfo{year}{1968}).

\bibitem{Karlstrom2003}
\bibinfo{author}{Karlstr\"{o}m, G.~{\it et al.}}
\newblock \bibinfo{title}{{MOLCAS: a program package for computational
  chemistry}}.
\newblock \emph{\bibinfo{journal}{Comp. Mat. Sci.}}
  \textbf{\bibinfo{volume}{28}}, \bibinfo{pages}{222} (\bibinfo{year}{2003}).

\bibitem{Chen2011}
\bibinfo{author}{Chen., K.~{\it et al.}}
\newblock \bibinfo{title}{{Molecular-ion trap-depletion spectroscopy of
  BaCl$^+$}}.
\newblock \emph{\bibinfo{journal}{Phys. Rev. A}} \textbf{\bibinfo{volume}{83}},
  \bibinfo{pages}{030501} (\bibinfo{year}{2011}).

\bibitem{Kielpinski2006}
\bibinfo{author}{Kielpinski, D.}, \bibinfo{author}{Cetina, M.},
  \bibinfo{author}{Cox, J.~a.} \& \bibinfo{author}{K\"{a}rtner, F.~X.}
\newblock \bibinfo{title}{{Laser cooling of trapped ytterbium ions with an
  ultraviolet diode laser.}}
\newblock \emph{\bibinfo{journal}{Opt. Lett.}}
  \textbf{\bibinfo{volume}{31}}, \bibinfo{pages}{757--9}
  (\bibinfo{year}{2006}).


\end{thebibliography}
\end{document}